\documentclass{emulateapj}
\usepackage{graphicx}
\usepackage{threeparttable}
\usepackage{hyperref}
\usepackage{color}

\newcommand{\xmm}{{\it XMM-Newton} }
\newcommand{\xmmp}{{\it XMM-Newton}}
\newcommand{\chandra}{{\it Chandra} }
\newcommand{\chandrap}{{\it Chandra}}
\newcommand{\ergs}{${\rm erg \ cm^{-2} \ s^{-1}}$ }
\newcommand{\erg}{${\rm erg \ s^{-1}}$ }

\begin{document}

\title{RX J1301.9+2747: A Highly Variable Seyfert Galaxy with Extremely Soft X-ray Emission}
\author{Luming Sun\altaffilmark{1}, Xinwen Shu\altaffilmark{1}, and Tinggui Wang\altaffilmark{1}}
\altaffiltext{1}{
CAS Key Laboratory for Research in Galaxies and Cosmology, Department of Astronomy, University of Science and Technology of China, Hefei, Anhui 230026; lmsun@mail.ustc.edu.cn, xwshu@mail.ustc.edu.cn, twang@ustc.edu.cn
}
\begin{abstract}
{ In this paper} we present a temporal and spectral analysis of X-ray data from { the} \xmm
and \chandra observations of the ultrasoft and variable Seyfert galaxy
RX J1301.9+2747. In both observations the source clearly { displays} two distinct
states in the X-ray { band}, a long quiescent state and a short flare (or eruptive) state { which} differs
in count rates by a factor of 5--7. The transition from quiescent to flare state
occurs in 1--2 ks. { We have observed} that the quiescent state spectrum
is unprecedentedly steep with a photon
index $\Gamma\sim7.1$, and { the spectrum of the flare state} is flatter with $\Gamma\sim4.4$.
X-rays above 2 keV { were} not significantly detected in either { state}.
In the quiescent state, the spectrum
appears to be dominated by a black body component of
temperature about $\sim$30--40 eV, which is { comparable to the expected}
maximum effective temperature from the inner
accretion disk. The { quiescent state however}, requires an additional steep power-law, presumably
arising from the Comptonization by transient heated electrons. Optical spectrum from the Sloan Digital Sky Survey { shows} Seyfert-like narrow
lines for RX J1301.9+2747, { while the HST imaging reveals a central point source for the object.}
In order to precisely determine the hard X-ray component,
future longer X-ray observations are required.
{ This will} help constrain the accretion disk model for RX J1301.9+2747, and shed new
light into the characteristics of the corona and accretion flows around black holes.

\end{abstract}


\section{Introduction}

Active galactic nuclei (AGNs) are thought to be powered by supermassive
black holes of $M_{\rm BH}\sim10^6-10^9~M_{\odot}$ accreting the surrounding gas \citep[see][for a review]{Rees84}.
{ They are also considered} to be scaled-up versions of Galactic black hole binaries (BHBs, $M_{\rm BH}\sim10~M_{\odot}$,
\cite{McHardy06} and references therein).
The rapid X-ray variability is one example of the similarities between these two types of systems
\citep{Gierlinski08}.
In Seyfert galaxies, variations of the X-ray continuum emission
over a timescale from minutes to
hours have been reported
\citep[e.g.][]{Ulrich97, Boller97, Ponti12},
{ however}, persistent giant and rapid variability
appears to be fairly rare and its origin is still poorly understood.

Soft X-ray excesses above an extrapolation of the
underlying hard X-ray power-law is commonly observed in Type 1 AGNs and radio quiet quasars \citep{Piconcelli05,Bianchi09}.
The origin of this additional component is not clear,
and may be the high-energy tail of the AGN accretion disk emission
\citep[e.g.][]{Grupe95}.
The problem { using} this explanation is that the temperatures of
the soft X-ray excesses appear to fall within a narrow range (kT $\sim$ 0.1--0.2 keV)
from a sample of AGNs { containing} a large range of black hole (BH) mass,
which is { difficult} to explain using the standard accretion disk models \citep[e.g.][]{Gierlinski04,Crummy06}.

So far there has { been} no convincing
evidence for the presence of direct accretion disk emission seen in the X-ray spectra of AGNs.
\cite{Yuan10} reported a luminous ultra-soft excess in the narrow line Seyfert 1 galaxy (NLS1) J1633+4718
from archival \textit{ROSAT} spectra,
and found a lowest soft excess temperature of 32 eV among AGNs.
{ This characteristic of the soft excess is }likely an
observational signature for the accretion disk emission.
However, the blackbody nature of this emission { needs} to be tested further, utilizing higher
quality X-ray data, as
the \textit{ROSAT} spectra ($\sim${ 0.1--2.4} keV) are less sensitive to constrain the harder power-law emission above
$\sim$2 keV.
Recently, \cite{Terashima12} reported the discovery of a candidate 'ultrasoft' AGN,
{ whose X-ray spectrum can be represented $purely$ by a soft thermal component with a
blackbody temperature of $kT\sim$0.13--0.15 keV},
by analog with the accretion disk dominated spectrum typically seen in the high/soft state of BHBs.
{ Additionally}, the soft X-ray emission obtained shows spectral variability consistent with being caused by strong Comptonization.
Interestingly, the object was later optically confirmed to be a Type 2 AGN \citep{Ho12}
with a central BH mass as small as $10^5M_{\odot}$. { However, in their work they
did not test in detail the possibility of the accretion disk emission as the origin for the soft excess.}

In this { paper}, we report results of new \xmm and \chandra observations of RX 1301.9+2747
at $z$=0.0237 (hereafter J1302), { a highly variable and ultra-soft} X-ray source
in a post-starburst galaxy \citep{Dewangan00}.
Our detailed analysis of the optical spectrum from the SDSS revealed that it is
{ a Seyfert galaxy}.
The ultrasoft X-ray emission of J1302 was confirmed in the new X-ray observations.
In particular, we found unusual giant flares in both \xmm and \chandra light curves,
accompanied by spectra hardening during the flare state.
{ In the quiescent state}, the spectrum appears to be dominated by a thermal blackbody component,
{ whose temperature is comparable to the predicted maximum accretion disk temperature.}
Throughout this paper,
we assume a cosmology with $H_0$ = 0.71, $\Omega_M$ = 0.27, $\Omega_\Lambda$ = 0.73.

\section{Data analysis and result}

\subsection{Optical Spectrum}
\label{2.1}

J1302 was spectroscopically observed by the SDSS in March 2007.
Figure 1 shows the rest-frame spectrum for J1302 (black line), which is dominated by the starlight of the host galaxy.
To subtract the starlight and the nuclear continuum, we followed the recipe described in detail in \cite{Dong05}.
As seen in Figure 1, the galaxy starlight model (green line) gives a very good fit to
the optical continuum ($\chi^2$/dof = 3648/3208).
After the subtraction of stellar absorption lines, we fitted the emission-line spectrum,
represented by a blue line, by
using Gaussians to derive the parameters of the emission lines.
{ [OIII] $\lambda\lambda$4959, 5007, [NII] $\lambda\lambda$6548, 6583, H$\alpha$ and [SII] $\lambda\lambda$6717, 6731}
emission lines are clearly detected with { S/N $>$ 5},
while H$\beta$ line is only weakly { distinguishable} with S/N $\sim$ 1.4.
The right panel in Fig. 1 { displays} the emission-line spectra, { alongside} the best fit Gaussian models.
There is no apparent broad component of H$\alpha$ line,
and { a} narrow Gaussian with a line width of FWHM $\sim$ 240km s$^{-1}$ can provide good fit.
In order to verify the absence of the broad H$\alpha$ line, we add an additional Gaussian to the narrow H$\alpha$,
with the width fixed at 2000 km s$^{-1}$. The flux is allowed to vary in the fitting process.
We found that the fitting was marginally improved by adding this component,
and the S/N for the H$\alpha$ broad component is only $\sim$0.5.
This suggests that the broad H$\alpha$ line, if there is any, is extremely weak in J1302.
The ratios of the narrow lines [OIII] $\lambda$5007/H$\beta\,>$ 4.8 (using 3$\sigma$ upper limit of H$\beta$ line flux),
and [NII] $\lambda$6583/H$\alpha$ = 2.3,
place J1302 into the { Seyfert regime} on the BPT diagram of \cite{Kewley06}.
{ The flux ratios of [SII]/H$\alpha$ and [OI]/H$\alpha$ are 0.56 and 0.32, respectively,
further strengthening the Seyfert nature of J1302 according to line ratio diagnostic diagrams of
\cite{Kewley06}.}

\subsection{X-ray Observations}
\label{2.2}

J1302 was observed by \xmm EPIC cameras in December 2000 with a total exposure time of 29 ks.
It was detected $\sim$7.3 arcmin away from the center of the field of view in the \xmm imaging of
the Coma cluster (ObsID 0124710801).
The \xmm data were reprocessed with the Science Analysis Software version 11.0.0,
using the calibration files as of December 2011.
We used principally the PN data, which have much higher sensitivity, using the MOS
data only to check for consistency.
Spectra and light curves of { source} were extracted from a circular region with a radius of
40$\arcsec$ centered at the source position for both PN and MOS cameras.
Background spectra were made from source-free areas on the same chip using four
circular regions { identical to} the source region.
The epochs of high background events were examined and excluded by using the
light curves in the energy band above 12 keV.

The \chandra pointing observation of J1302 was taken in June 2009 for about 5 ks.
The data were processed with \texttt{CIAO} (version 4.3) and CALDB (version 4.4.1), following the
standard criteria.
Fig. 2 shows the \chandra X-ray contours of J1302, overlaid on the { HST image in the B band}.
It is clear from the figure that the X-ray emission of J1302 is point-like.
The center for the X-ray source is coincident with the optical nucleus of the galaxy
(with a positional offset of $\sim$0.1$\arcsec$).
Given the subarcsecond spatial resolution of \chandrap, we conclude that most of the X-ray emission,
{ if not all,}
comes from the nuclear region of the galaxy, likely related to the AGN.

\subsection{X-ray light curve}
\label{2.3}

Light curves from the \xmm and \chandra observations are shown in Fig. 3.
The source exhibits large-amplitude count rate variations in both observations.
It can be seen from the PN light curve that there is a giant flare with count rates rising
by a factor of 5 times the average value, { having} a duration of $\sim$2 ks.
The X-ray flux then declines to a relatively steady state. Such X-ray flare is confirmed by the MOS light curves,
in which a possible decline of another flare is also recorded { at the beginning of the MOS observation}.
The time interval of the two flares is about 17 ks.
Interestingly, similar flare is seen in the \chandra light curve, with count rates increasing by a factor of 7 within $\sim$1 ks.
The similar amplitudes of flares in the \xmm and \chandra observations which span $\sim$ 9 years
suggest that the flare behaviour in J1302 seems persistent on time scale of $\sim$ decade.
The spectral variability during the flare will be explored in detail in Section~\ref{2.4}.

\subsection{X-ray Spectra}
\label{2.4}

As both \xmm and \chandra observations show peculiar flare behaviours,
we attempt to quantify the spectral variability during flares by dividing the data into high and low flux intervals,
using count rate thresholds of 0.35 counts s$^{-1}$ for \textit{XMM}-PN and 0.08 counts s$^{-1}$
for \chandrap, respectively.
{ For simplicity, we classify the data above the count rate thresholds as belonging
to the flare state, and that which falls below, to the quiescent state.}
The spectra data were grouped { in the following manner}: data from the \xmm
had at least 20 counts per bin to ensure the $\chi^2$ statistics, and
the \chandra data had at least 3 counts per bin and utilized
the $C$-statistics which was adopted for minimization.
Spectral fitting was performed using the \texttt{XSPEC} (Version 12.6) and
limited to the 0.2--2 keV range for \xmmp, since the emission is background dominated above that energy range.
The \chandra data was fitted in the energy range between 0.3 and 3 keV.
{ Throughout} the model fittings,
the Galactic column density was considered and
fixed at $N_{\rm H}^{\rm Gal}=0.75\times10^{20}$ cm$^{-2}$ \citep{Kalberla05}.

Spectral variability is clearly present between the two states with the source
being much harder in the flare state.
To illustrate the differences in the spectral slope between the two states,
we show the spectra together with an absorbed power-law model in Figure 4.
While the fit is generally acceptable for the two states, as confirmed
by the reduced $\chi^2$ value (see Table 1),
the photon indices obtained from a power-law fit are, however, extremely steep with
{ $\Gamma$ = $4.4^{+0.5}_{-0.4}$ for $N_H=4.3^{+2.2}_{-1.8}\times10^{20}$ cm$^{-2}$, and $\Gamma$ = $7.1^{+0.9}_{-0.7}$
for $N_H=3.6^{+1.9}_{-1.6}\times10^{20}$ cm$^{-2}$}, for the \textit{XMM} { flare and quiescent} state, respectively.
The photon indices belong to the steepest values obtained from
AGN X-ray spectra. For comparison,
the mean photon index in the 0.2--2.0 keV band
is $\sim2.9$ for a sample of soft X-ray selected AGNs observed with the \textit{ROSAT} RASS \citep{Grupe10}.
{ The result of this comparison indicates that} the X-ray spectrum of J1302 is extremely soft
{ compared to other AGNs.}
The absorption-corrected luminosity in the 0.5--2.0 keV range for { this} simple power-law model
{ is $6.7\times10^{41}$ and $2.8\times10^{40}$ \erg} for the \textit{XMM} { flare and quiescent} state, respectively.

In order to further investigate the spectral variability in J1302,
we { then} attempted to fit the spectra with a blackbody (\texttt{bbody} in \texttt{XSPEC}) or Multiple
Color Disk model (MCD, \texttt{diskbb} in \texttt{XSPEC}),
and a thermally Comptonized disk model \citep[\texttt{compTT} in \texttt{XSPEC},][]{Titarchuk94},
{ both of these alternative models have been used to fit the spectra of Galactic BHBs \citep[e.g.,][]{Done07},
and the soft X-ray excess emission in AGNs \citep[e.g.,][]{Porquet04,Patrick12} }.
{ The \texttt{diskbb} model integrates over the surface of accretion disk to form a multicolor
blackbody spectrum, and \texttt{compTT} is an analytic model that self-consistently
calculates the spectrum produced by the
Comptonization of soft seed photons in a hot corona above the accretion disk.
The physical parameters of the \texttt{compTT} model are: the soft photon temperature ($kT_0$), the temperature of the
Comptonizing electrons ($kT_e$), the plasma scattering optical depth ($\tau$).
For our fitting with the \texttt{compTT} model, a disk geometry
was assumed for the comptonizing region, and
the seed photons were assumed to follow Wien's law with a temperature of 22 eV
(the expected disk temperature in section~\ref{3.2}).
Because the temperature and optical depth of the Comptonizing plasma are strongly coupled
(both are equally involved in shaping the spectrum) and thus cannot be constrained simultaneously,
we fixed the plasma temperature at 20 keV and obtained constraints on the optical depth\footnote{
Leaving the plasma temperature as a free parameter
yields $kT_e=21(<27)$ keV and $\tau<1.3$ for the XMM quiescent state spectrum, but both parameters
cannot be constrained by the data during the flare.}.
The single Comptonized model yields consistent fitting results with the previous simple power-law model for
the spectrum at both states.
The Compton optical depth is
$\tau=0.16^{+0.07}_{-0.05}$ and $\tau<0.03$ for the flare and
quiescent state, respectively.
%
For the spectrum in the \xmm { flare state}, a multicolor-disk blackbody gives equivalent fit,
which is statistically better than the simple blackbody emission.}

{ The spectrum at the \xmm quiescent state, however, shows an excess of emission at energies above $\sim$ 0.7 keV
when fitted with a thermal model (\texttt{bbody} or \texttt{diskbb}).
The addition of a power-law to the model improves the fit with very high statistical significance
($\chi^2$ decreased by 13.9 for two extra parameters, at a 99.98\% level according to $F$-test).
The power-law component contributes $\sim$15\% of the total luminosity in the 0.3--2 keV band.
In this case, we obtain an effective blackbody temperature of $kT_{\rm BB}=43^{+6}_{-3}$ eV,
comparable to the seed photon temperature assumed in the \texttt{compTT} model. }
Although with large uncertainties,
the additional power-law component is relatively steep with photon index $\Gamma=4.3^{+1.6}_{-1.9}$, and
it is close to what is observed in the { flare state}.
The spectral fitting results for the PN data, alongside the observed flux and intrinsic
luminosity in the 0.5--2 keV range for each model, are shown in Table 1.
Note that we used the same models to fit the MOS data and found that the results agree well with the PN data.

The \chandra spectra were fitted with the same models used in the \xmm observation { and the results are listed in Table 1}.
{ During the first run we found that
the photon indices derived from the simplest power-law model are systematically
flatter than the values for the \xmm data.
The spectrum during the flare can be well fitted by a power-law model ($C$/dof=22.5/22).
On the other hand, a power-law model is not sufficient to fit the data at
the \chandra quiescent state.
%
%
The addition of a soft thermal component with respect to the
power-law model improves the fit significantly ($C$ value decreases by $\sim$9
for two extra parameters, corresponding to a significance level of 98.7\%).
The resulting best-fit photon index is flatter, with $\Gamma=3.5^{+0.8}_{-1.0}$.
Similarly, we obtain an effective disk temperature,
when fitted with a blackbody, of $kT_{\rm BB}=29^{+19}_{-16}$ eV.
This value is slightly lower than the blackbody temperature for the \xmm quiescent state,
but the parameter is loosely constrained due to the poor statistics of the \chandra data.
The power-law component in this case contributes $\sim40\%$ of the total luminosity in the 0.3--2 keV,
indicating a possible change of the power-law or the blackbody emission between \chandra
and \xmm observations.

Although a Comptonized model as opposed to a power-law model also provides a good fit
for the \chandra flare state spectrum, it is not sufficient to fit the data at the quiescent state.
The addition of an extra hard power-law component is needed at a significance level of 98.7\%
according to $F$-test.
%
The best-fitted optical depth $\tau$ for the \texttt{compTT} model is
not significantly different from that obtained with the \xmm data. }
{ The unabsorbed luminosity in the 0.5--2 keV band,
based on the best-fit power-law model and the \texttt{diskbb}+power-law model
for the \chandra { flare and quiescent state}, is $5.1\times10^{41}$ and $4.4\times10^{40}$ erg s$^{-1}$, respectively.}
%

\section{Discussion}

\subsection{AGN characteristics in RX J1301.9+2747}
\label{3.1}

X-ray observation with \chandra, which has superb spatial resolution of $\sim0.5\arcsec$,
revealed the presence of an AGN in J1302: the center of the bright unresolved X-ray
emission coinciding with the optical point-like nucleus of the galaxy (with a position offset of $\sim0.1\arcsec$).
Both the \chandra and \xmm observations show that it has Seyfert-like X-ray luminosity
of $\sim10^{41}$ erg s$^{-1}$ in the energy range of 0.5--2 keV ({ $\sim10^{42}$ erg s$^{-1}$
in the 0.2--2 keV band }),
and rapid X-ray variability down to a time-scale $\sim$ 1 ks.
Additionally, the optical spectrum { of J1302 displays Seyfert-like} narrow emission line ratios.
All these observational facts point to the presence of an AGN (Seyfert nucleus) in this galaxy.
This can be further supported by the point-like appearance of a nuclear source
from the HST imaging observations with $\sim0.1\arcsec$ resolution \citep{Caldwell99}.

Based on \textit{ROSAT} X-ray observations, \cite{Dewangan00} argued for the presence of an AGN in J1302,
{ a view that is consistent with ours on the basis of the new observations}.
However, based on their observed optical spectrum, Dewangan et al.
conclude that the galaxy nucleus is more like a LINER.
In this paper, we have carefully modeled the host galaxy's starlight, especially the
stellar absorption features, and subtracted them from the new SDSS spectrum,
enabling us to accurately measure the weak AGN emission lines
in J1302 \citep[e.g.,][]{Dong05},
thus more { resolutely confirming} the Seyfert nature based on the line ratio diagnostics.

The lack of detectable broad permitted lines prevents us from
estimating the central BH mass of J1302 using
conventional linewidth-luminosity-mass scaling relation.
In the SDSS spectral fitting, we found
the strongest narrow line, [OIII]$\lambda$5007, is marginally resolved with Gaussian $\sigma=58\pm9$ km s$^{-1}$
(after correcting for the instrument resolution).
Using the width of the [OIII]$\lambda$5007 line as a proxy for the stellar velocity dispersion of the host galaxy,
we obtained a BH mass of $M_{\rm BH} = 8\times10^5M_{\odot} $ with an intrinsic
scatter of 0.5 dex \citep[e.g.,][]{Xiao11}.

The bolometric luminosity for J1302 can be estimated from the
optical continuum luminosity.
{ We retrieved the high resolution HST/WFPC2 images of J1302 in the
$B$ (F450W filter) and $I$ (F814W filter) passbands from the HST archive.
The HST observations (dataset U39D0301M--U39D0304M) were made in July 1997,
with two 600 s exposures in the $B$ band and two 400 s exposures in the $I$ band, respectively.
The images were processed using the standard HST pipeline routines in IRAF/STSDAS\footnote
{http://www.stsci.edu/institute/software\_hardware/stsdas}.
We then performed two-dimensional profile decompositions of this galaxy
with the code \texttt{GALFIT} (version 3.0, Peng et al. 2010).
Our model consists of an exponential disk component, a bulge component,
and an unresolved central point source for
the nuclear AGN emission. In our \texttt{GALFIT} modeling, the point-spread function
was generated by the $\tt{Tiny\,Tim}$ software \citep{Krist95}.
Note that the bulge for this galaxy displays a box/peanut shape, which is commonly seen in edge-on barred disk galaxy,
{consequently we added a boxiness parameter to the bulge profile in \texttt{GALFIT}.
The model generally matches the data well ($\chi^2_\nu\sim1.02$).}
The inferred flux for the central point source in the $B$-band is $M_B$ = -15.8,
corresponding to a nuclear luminosity of $\nu L_{\nu B}$ = $8\times10^{41}$ erg s$^{-1}$.
For comparison, the B-band bulge luminosity of this galaxy derived from the \texttt{GALFIT} decomposition is
{ L$_{\rm B, bulge}$} $\sim6\times10^{42}$ erg s$^{-1}$}.
If indeed the nuclear emission comes from AGN\footnote{Note that we have not accounted for the
dust extinction in estimating the luminosity.},
we estimate the bolometric luminosity to be $1\times10^{43}$ \erg { using the $B$-band
luminosity for the central point source} by adopting a bolometric correction of 13 \citep{Marconi04}.
For a BH mass of $8\times10^5 M_\odot$, the accretion rate in Eddington unit is $L/L_{\rm{EDD}} \sim$ 0.1,
{ which suggests} J1302 is accreting at high Eddington ratio.

\subsection{Extremely Soft X-ray Emission}
\label{3.2}

One of the remarkable features of J1302 is the extreme softness of the X-ray spectra.
The best-fitted power-law index for the spectrum in the { \xmm quiescent state}
($\Gamma\sim7$) is one of the steepest soft X-ray photon indices among AGNs \citep[e.g.,][]{Grupe95, Boller11}.
Understanding the origin of the ultrasoft X-ray emission will help
to pin down the nature of the source.
Some AGNs such as NLS1s can be very soft \citep[e.g.,][]{Boller96,Middleton07},
showing a strong soft X-ray excess
over an underlying power-law component.
The strength of soft excess can be quantitatively
described as the ratio { of flux }at 0.5 keV to the power law extrapolation of the fitting to the spectrum
above 2 keV \citep{Middleton07}.
Since no significant hard X-ray emission above $\sim$2 keV was detected in the
\textit{XMM} observation of J1302, we estimated 90\% confidence upper limit on the count rates
in 2--7 keV, $\sim6.7\times10^{-4}$ counts s$^{-1}$,
and converted it to an upper limit on the extrapolated flux at 0.5 keV.
\texttt{XSPEC} simulations of models with power-law $\Gamma$ = 2, 2.5 and 3 show that
the lower limits on the ratios defined above are 37, 15 and 6.1, respectively.
Note that the ratios for a sample of NLS1s \citep[]{Middleton07}
are usually found to be less than 10.
Thus, the soft emission relative to that at { the} hard X-rays in J1302
is extremely strong { compared to other AGNs}.


The origin of soft excess is still unclear.
The spectrum at the \textit{XMM} { quiescent state} can be fitted well by
a Comptonized model { and a blackbody plus power-law emission equally.
Interestingly}, the fitted blackbody temperature ($kT_{\rm BB}=43^{+6}_{-3}$ eV)
is much lower than the canonical values
of $\sim$ 0.1--0.2 keV found for AGNs \citep{Crummy06}.
Standard accretion disk models \citep{Shakura73}
give a maximum effective temperature of the accreting material
$kT_{\rm max}\sim$ $11.5(\dot{m}/M_8)^{1/4}$ eV, where $\dot{m}$ is mass accretion rate
in Eddington unit and $M_8$ = $M_{\rm BH}/10^8M_\odot$.
Using $M_{\rm BH}=8\times10^5 M_\odot$ and $\dot{m}\sim0.1$ for J1302, we
obtained $kT_{\rm max}\sim$22 eV, which is comparable to the fitted blackbody temperature.
Note that though with larger uncertainties, the lower fitted temperature for the
\chandra { quiescent state} data ($kT_{\rm BB}=29^{+19}_{-16}$ eV) is more compatible with the predicted maximum disk temperature.
Therefore, the ultrasoft X-ray emission in J1302 may be connected with the
direct thermal emission from the accretion disk.
Another constraint on the X-ray spectra due to thermal disk emission can come from the optical/UV data.
The optical $B$-band flux of nuclear point source from the HST observation is,
however, much higher than the extrapolated MCD flux in the optical ($M_B\sim-12.7$),
about one order of magnitude difference.
The difference cannot be explained by the contamination from nuclear star clusters,
as they have typical absolute $I$-band magnitudes between $-14$ and $-10$ \citep{Boker02},
much lower than the observed $I$-band flux of J1302 nucleus ($M_I\sim-16.8$,
{ obtained with the same GALFIT decomposition of the HST/WFPC2 image as detailed in Section~\ref{3.1})}.
To further investigate this difference we need a more realistic disk spectral
modeling, and fit to a broader band data,
which is beyond the scope of this paper.

\subsection{Unusual X-ray variability}
\label{3.3}

Another unusual feature of J1302 is that it clearly shows two distinct states in the X-ray,
a flare (or eruptive) state and a quiescent state.
The amplitudes of the flare in the \chandra and \xmm light curves look very similar,
with the count rates increased by a factor of 5--7 within $\sim$1--2 ks.
Both the \xmm observation in 2000 and the \chandra observation in 2009 detected rapid
flares, suggesting that the flare itself appears repetitive and occurs very
frequently in the object.
In fact, {\it ROSAT} PSPC observations also found that the object is highly
variable and { demonstrates a rapid flare event in light curve lasting $\sim$1.3 ks} \citep{Dewangan00}.
The rapid energetic flare in J1302 is fairly rare among Seyfert galaxies and quasars,
although some extreme variations { have} also been found in objects such as
IRAS 13224-3809 \citep{Boller97} and PKS 0558-504 \citep{Wang01}.
As discussed by \cite{Wang01}, a magnetic coronal model in which electrons
in the corona are continuously heated by magnetic reconnection, can produce rapid energetic X-ray
{ flare}.
A realistic physical model to explain a rapid energetic flare in AGNs, and the
comparison with { Galactic BHBs} are beyond the scope of this work and will
be presented elsewhere.

The spectral variability is clearly present between the two flux states with
the source being much harder in the { flare state}.
This behavior is markedly contrast to what is commonly seen in AGNs and BHBs
\citep[e.g.,][]{Markowitz04, Done07}.
Clearly the spectral variability cannot be explained by any simple spectral model.
One possibility is that the spectrum in the { quiescent state}
is represented by a relatively stable thermal emission from accretion disk,
and the spectral variability is caused by the flux variations of
a second additional component
such as strongly Comptonized power-law emission.
This perhaps explains why the photon index for the power-law emission
in the \xmm { flare state} (which dominates the spectrum) is consistent
with the { additional} power-law component in the { quiescent state},
whose strength is relatively weak.
In this case, the underlying power-law emission is still steep ($\Gamma\sim$4)
compared to other AGNs, the nature of which remains understood.
{ Note that similar spectral change has also been seen in the \chandra observation of J1302,
though the spectra statistics for the data is low.   }
Longer X-ray observations are required to examine the presence of persistent flares and to
investigate the origin of spectral variability during flare periods.
This can in turn shed new lights on the characteristics of corona and accretion flows around BHs.

\begin{acknowledgements}
This research made use of the HEASARC online data archive
services, supported by NASA/GSFC. { We would like to thank the anonymous
referee for his/her helpful comments to improve the content of the paper.
We are also grateful to G. Mckenzie for offering language assistance}.
This work was supported
by Chinese NSF through grant 11103017, 11233002,
and National Basic Research Program 2013CB834905.
{ X.S. acknowledges the support from the Fundamental Research Funds for the Central
Universities (grant Nos. WK2030220004, WK2030220005)}.
\end{acknowledgements}

\begin{table}[!t]\footnotesize
\caption{Spectral fitting results for the \xmm and \chandra observation at different states}
\begin{threeparttable}
\begin{tabular}{ccccccccc}
\hline
\hline
\multicolumn{9}{c}{\textit{XMM}-PN flare state}\\
\hline
wabs*model\tnote{a}&$\rm{N_H}$     &$\Gamma$            &kT          &$\tau$\tnote{b} &$\chi^2$/d.o.f. &F$_{0.5-2\rm{keV}}$\tnote{c} &L$_{0.5-2\rm{keV}}$\tnote{c}&\\
&($10^{20}\,\rm{cm}^{-2}$)&&(eV)&& &($10^{-14}$ erg s$^{-1}$ cm$^{-2}$) &($10^{40}$ erg s$^{-1}$)&\\
\hline
powl          &$4.3^{+2.2}_{-1.8}$ &$4.4^{+0.5}_{-0.4}$ &                  &                       &40.4/28 &40&67&\\
bbody         &0.75(fixed)\tnote{d}&                    &$99^{+6}_{-5}$    &                       &55.5/29 &38&54&\\
diskbb        &0.75(fixed)         &                    &$134^{+9}_{-9}$   &                       &43.7/29 &39&55&\\
compTT\tnote{e}&$4.3^{+2.2}_{-1.8}$&                    &                  &$0.16^{+0.07}_{-0.05}$ &40.3/28 &40&67&\\
\hline
\multicolumn{9}{c}{\textit{XMM}-PN quiescent state}\\
\hline
powl          &$3.6^{+1.9}_{-1.6}$ &$7.1^{+0.9}_{-0.7}$ &                 &                      &27.9/40 &1.6&2.8&\\
bbody+powl    &0.75(fixed)         &$4.3^{+1.6}_{-1.9}$ &$43^{+6}_{-3}$   &                      &26.3/39 &1.7&2.5&\\
diskbb+powl   &0.75(fixed)         &$3.7^{+2.0}_{-2.0}$ &$52^{+5}_{-5}$   &                      &25.9/39 &1.7&2.6&\\
compTT        &$2.9^{+1.4}_{-1.3}$ &                    &                 &0.016($<$0.03)        &27.7/40 &1.6&2.7&\\
\hline
\hline
\multicolumn{9}{c}{\chandra flare state}\\
\hline
wabs*model    &$\rm{N_H}$          &$\Gamma$            &kT              &$\tau$ &$C$/d.o.f. &f$_{0.5-2\rm{keV}}$ &L$_{intr,0.5-2\rm{keV}}$&\\
&($10^{20}\,\rm{cm}^{-2}$)&&(eV)&&& ($10^{-14}$ erg s$^{-1}$ cm$^{-2}$) &($10^{40}$ erg s$^{-1}$)&\\
\hline
powl          &3.7($<$11)          &$3.2^{+1.1}_{-0.6}$ &                 &                       &22.5/22 &34&51&\\
bbody         &0.75(fixed)         &                    &$195^{+24}_{-20}$&                       &27.6/23 &35&47&\\
diskbb        &0.75(fixed)         &                    &$275^{+51}_{-38}$&                       &22.5/23 &35&47&\\
compTT        &3($<$20)            &                    &                 &$0.44^{+0.28}_{-0.28}$ &22.5/22 &34&50&\\
\hline
\multicolumn{9}{c}{\chandra quiescent state}\\
\hline
powl          &0.75(fixed)         &$4.5^{+0.6}_{-0.6}$ &                 &                       &16.5/14 &3.7 &5.3&\\
bbody+powl    &0.75(fixed)         &$3.5^{+0.8}_{-1.0}$ &$29^{+19}_{-16}$ &                       &8.0/12  &3.2 &4.5&\\
diskbb+powl   &0.75(fixed)         &$3.5^{+0.8}_{-1.0}$ &$32^{+25}_{-18}$ &                       &8.0/12  &3.2 &4.4&\\
compTT+powl   &0.75(fixed)         &2.8($<$3.9)         &                 &0.01($<$0.09)\tnote{f} &8.1/12  &3.1 & 4.3&\\
\hline
\hline
\end{tabular}
\begin{tablenotes}
    \item [a] Spectral model (as given in \texttt{XSPEC}) multiplied by a neutral absorption
(\texttt{wabs}) with column density $\rm{N_H}$. { \texttt{wabs} is the photo-electric absorption model using
Wisconsin--Morrison \& McCammon (1983) cross-sections.}
    \item [b] Plasma optical depth in the \texttt{CompTT} model.
    \item [c] { $F_{\rm 0.5-2 keV}$ is the observed 0.5--2 keV flux in units of
$10^{-14}$ \ergs. { $L_{\rm 0.5-2 keV}$ is the \textit{unabsorbed} luminosity in the energy range of
0.5--2 keV, in units of $10^{40}$ \erg.}}
    \item [d] The column density was fixed to Galactic value $\rm{N_H^{Gal}}$ if the fitting yields a $\rm{N_H}<$$\rm{N_H^{Gal}}$.
    \item [e] The spectrum of the seed photons is assumed to be Wien law with a temperature of 22 eV (see Section~\ref{3.2} for details). We fixed the plasma temperature at 20 keV and obtained constraints on the optical depth.
    \item [f] { Pegged at the minimum value allowed in \texttt{XSPEC}.}
\end{tablenotes}
\end{threeparttable}
\label{tab1}
\end{table}
\newpage
\clearpage

\begin{figure}
\centering{
 \includegraphics[scale=0.9]{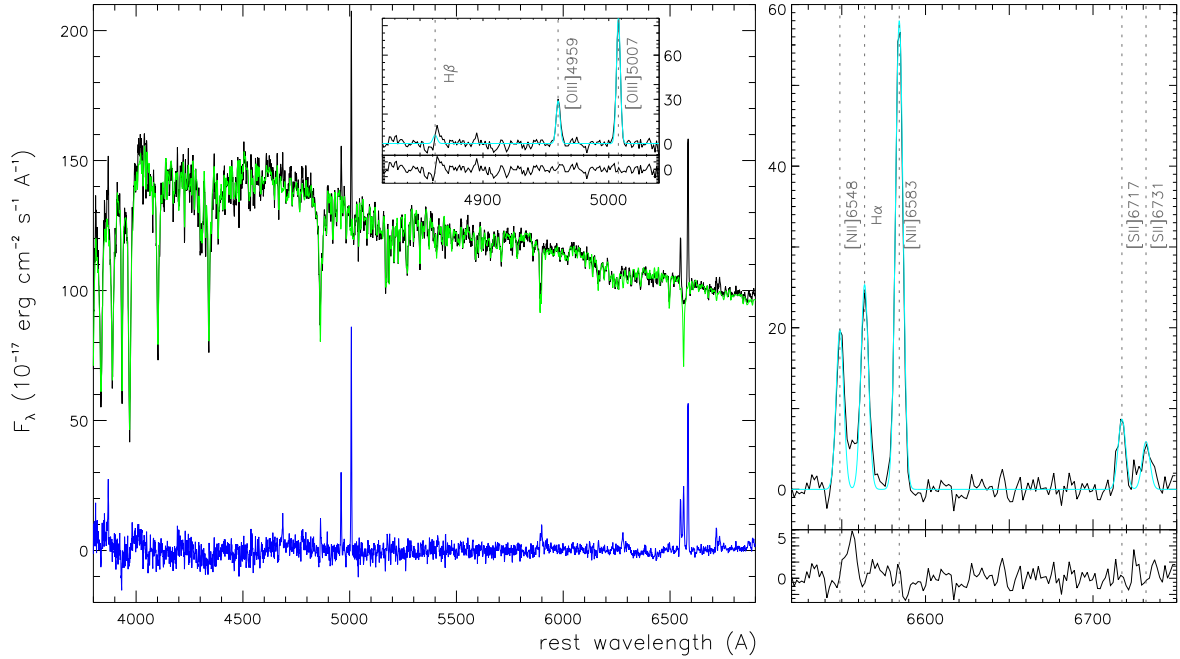}
 \caption{Illustration of the continuum and emission-line fittings of the SDSS spectrum.
{\it Left panel}: observed spectrum (black), stellar continuum model (green) and residual (blue) which is used to fit the emission lines.
{\it Right panel}: { a zoomed-in view of the emission-line profile fitting for the H$\alpha$+[NII] and the [SII] doublet lines}.
  The inset in the left panel shows a zoomed-in view of the H$\beta$ and [OIII] region.
  Gaussian line models are plotted in cyan line, { and the residual is shown in the lower panel}.
{ In the emission-line fits, the line ratios of [OIII] and [NII] doublet lines are fixed to
the theoretical values}.
}}
\label{SDSSspec}
\end{figure}

\begin{figure}
\centering{
 \includegraphics[scale=0.3]{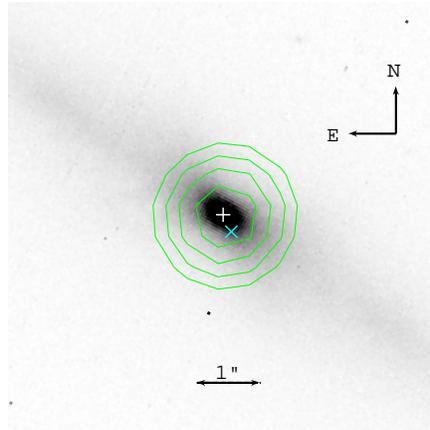}
 \caption{Contours of the \chandra image (green) overlaid with the { HST $B$-band} image.
{ The direction of north is up and east is left.
White plus marks the { position of the central point source from the HST imaging}.
For comparison, the center of the radio emission obtained with the Very Large Array \citep{Miller09} is shown in cyan cross.}}
}
\label{image}
\end{figure}

\begin{figure}
\centering{
 \includegraphics[scale=0.5]{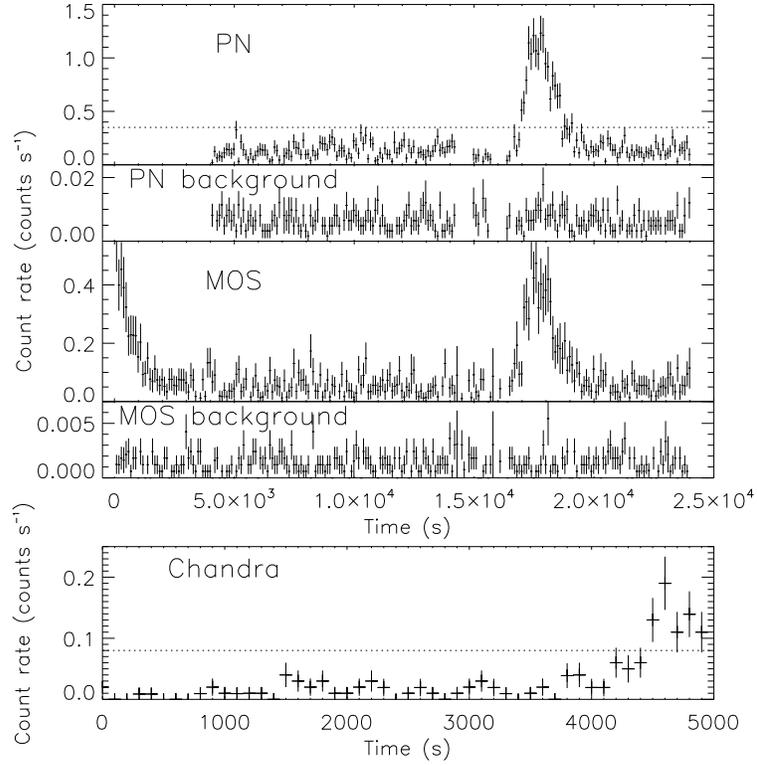}
 \caption{{ \xmm and \chandra light curves for J1302}.
{\it Top panel}: \textit{XMM}-PN and the summed MOS1+MOS2 light curves,
with { a time} bin size of 100 s.
The PN and MOS background light curves are also shown for comparison (lower panel).
{\it Bottom panel}: \chandra light curve with the same bin size as \xmmp.
{ The dotted lines represent the count rate thresholds dividing the data into flare and quiescent state, which
are 0.35 counts s$^{-1}$ for \textit{XMM}-PN, and 0.08 counts s$^{-1}$ for \chandra, respectively.}
}}
\label{lightcurve}
\end{figure}

\begin{figure}
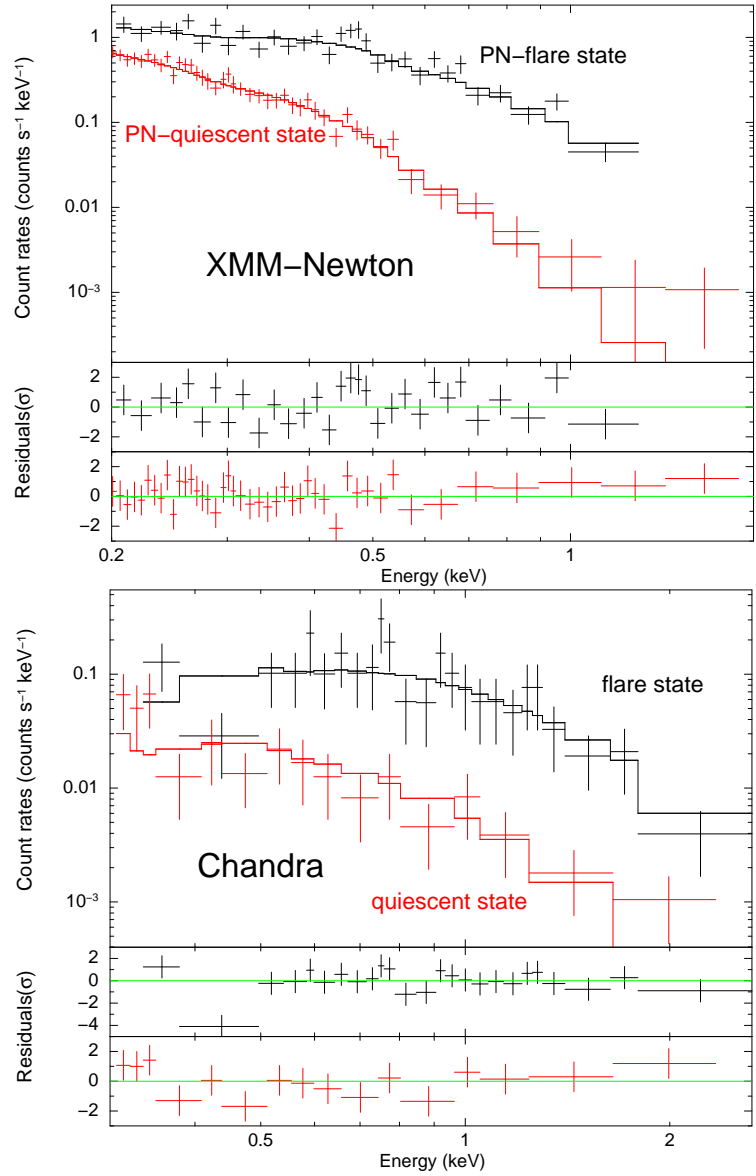

\centering{
 \includegraphics[scale=0.4,angle=270]{fig4a.ps}
 \includegraphics[scale=0.4,angle=270]{fig4b.ps}
 \caption{{\it Top panel}: Spectra of the \xmm flare state (black) and the quiescent state (red)
for J1302. Only the PN data are shown for clarity.
{ The solid lines are the simple power-law model fits for both states,
and the corresponding residuals are shown in the lower panels.}
{\it Bottom panel}: as top panel, but showing for the \chandra data.
}}
\label{Xrayspec}
\end{figure}

\end{document}